# Temperature-driven changes in the Fermi surface of graphite

Laxman R. Thoutam[1,2], Samuel E. Pate[1,3], Tingting Wang[4], Yong-Lei Wang[4,*], Ralu Divan[5], Ivar Martin[1], Adina Luican-Mayer[6], Ulrich Welp[1], Wai-Kwong Kwok[1], and Zhi-Li Xiao[1,3,*]

[1]*Materials Science Division, Argonne National Laboratory, Argonne, Illinois 60439, USA*

[2]*Department of Electronics and Communications Engineering, SR University, Warangal, Telangana 506371, India*

[3]*Department of Physics, Northern Illinois University, DeKalb, Illinois 60115, USA*

[4]*Research Institute of Superconductor Electronics, School of Electronic Science and Engineering, Nanjing University, Nanjing 210093, China*

[5]*Center for Nanoscale Materials, Argonne National Laboratory, Argonne, Illinois 60439, USA*

[6]*Department of Physics, University of Ottawa, Ottawa, Ontario K1N 6N5, Canada*

[*]Correspondence to: yongleiwang@nju.edu.cn; xiao@anl.gov

We report on temperature-dependent size and anisotropy of the Fermi pockets in graphite revealed by magnetotransport measurements. The magnetoresistances obtained in fields along the *c*-axis obey an extended Kohler's rule, with the carrier density following prediction of a temperature-dependent Fermi energy, indicating a change in the Fermi pocket size with temperature. The angle-dependent magnetoresistivities at a given temperature exhibit a scaling behavior. The scaling factor that reflects the anisotropy of the Fermi surface is also found to vary with temperature. Our results demonstrate that temperature-driven changes in Fermi surface can be ubiquitous and need to be considered in understanding the temperature-dependent carrier density and magnetoresistance anisotropy in semimetals.

## I. INTRODUCTION

The Fermi surface, defined in reciprocal space as the surface of Fermi energy $E_F$ separating the occupied electron states from unoccupied ones at zero temperature, is at the very heart of our understanding of the electronic properties of metallic states [1]. For example, anisotropic magnetoresistances of materials can arise from non-spheric Fermi surfaces [2-5]. Changes in the Fermi surface topology, i.e., the Lifshitz transition, can evoke superconductivity in a semimetal under pressure [6]. In the absence of a structural symmetry breaking and/or a magnetic phase transition, the Fermi surface in a conventional metal does not change appreciably with temperature $T$, since $E_F$ is typically much larger than $k_B T$ [7]. On the other hand, one of the interesting phenomena revealed in recent research is the temperature-driven Lifshitz transition [7-27], indicative of a temperature-induced reconstruction of the Fermi surface, discovered in the type-II Weyl semimetal $WTe_2$ [7]. Here, we report on the temperature dependence of the Fermi surface in the *absence* of a Lifshitz transition in graphite and demonstrate that temperature-driven changes in the Fermi surface may be ubiquitous in a semimetal, independent of a Lifshitz transition.

While angle-resolved photoemission spectroscopy (ARPES) remains a powerful technique in uncovering a temperature-induced Lifshitz transition [8-13], the conventional magnetotransport approach [15-27] is another method to study Lifshitz transitions [18-20,23-27]. As demonstrated recently in the nodal line semimetal ZrSiSe [18], a Lifshitz transition can be inferred from the anomalies in the temperature dependence of the carrier density $n(T)$ and/or mobility $\mu(T)$ by analyzing the measured magnetoresistance using a two-band model.

Here, we conduct magnetotransport investigations on semimetal graphite which exhibits no Lifshitz transition, i.e., without anomalies in $n(T)$ and/or $\mu(T)$. We chose graphite for the following reasons: (1) its low carrier density ($\sim 10^{18}$ cm$^{-3}$) [28-31] allows temperature-induced change in $E_F$ to be discerned from the temperature dependence of the carrier density [28]; (2) graphite is a two-band system, with the possibility to avoid the complexities associated with typical multi-band semimetals. Our results show that temperature-driven changes in the Fermi surface can be reflected in the temperature dependence of the carrier density [28] and magnetoresistance anisotropy [32]. In particular, magnetotransport results reveal both the temperature-induced shift of the Fermi level and change in the anisotropy of the Fermi surface in graphite. Our results demonstrate that temperature-induced changes in the Fermi surface may be expected in semimetals without a Lifshitz transition. They also indicate that temperature-induced changes in the Fermi surface need to be considered in understanding the temperature behavior of a semimetal, such as its temperature-dependent carrier density and the anisotropy of the magnetoresistances.

We first investigate the magnetoresistances obtained in magnetic fields along the $c$-axis of the crystal and at various temperatures. We find that they follow an extended Kohler's rule [33]:

$$MR = f[H/(n_T \rho_0)] \qquad (1)$$

where $MR = [\rho_{xx}(H) - \rho_0]/\rho_0$ is the magnetoresistance, $n_T$ is the thermal factor representing the temperature dependence of the carrier density, $\rho_{xx}(H)$ and $\rho_0$ are the longitudinal resistivities at a magnetic field $H$ and zero field at a given temperature, respectively. The good scaling of the data to the extended Kohler's rule suggests that the size of the graphite's Fermi pockets changes with temperature. We further probe the angle dependence of the magnetoresistivities at a given temperature and find it can be scaled as [32]

$$\rho_{xx}(H, \theta) = \rho_{xx}(\varepsilon_\theta H) \qquad (2)$$

where $\varepsilon_\theta = (cos^2\theta + sin^2\theta/\gamma^2)^{1/2}$, with $\theta$ being the magnetic field angle with respect to $c$-axis of the crystal. The scaling factor $\gamma$ varies from $\gamma = 56$ at $T = 2$ K to $\gamma = 20$ at $T = 300$ K. Since $\gamma$ is associated with the anisotropy of the effective mass [32], its change with temperature evinces a temperature-dependent anisotropy of the Fermi surface.

## II. MATERIALS AND METHODS

We measured samples that were mechanically exfoliated out of a natural graphite crystal purchased from NGS Trading & Consulting GmbH, Germany [34]. Electric contacts with well-defined separations and locations were achieved using photolithography followed by evaporation deposition of 300-500 nm thick Au layer with a 5 nm thick Ti adhesion layer (see Fig.S1 for an image of the sample from which the reported data were obtained). DC four-probe resistive measurements were carried out in a Quantum Design PPMS-9 using a constant current mode ($I$ = 100 μA). Angular dependencies of the resistance were obtained by placing the sample on a precision, stepper-controlled rotator with an angular resolution of 0.05°. The magnetic field is always perpendicular to the current $I$ which flows in the $ab$ plane of the crystal.

## III. RESULTS AND DISCUSSION

Figure 1(a) (in logarithmic scale) and Fig.S2 (in linear scale) present the typical magnetic field dependence of the longitudinal resistivity $\rho_{xx}(H)$ at various temperatures. At $T$ = 300 K, $\rho_{xx}(H)$ can be well described as $\rho_{xx}(H) \sim H^\alpha$ with $\alpha = 1.8$. With decreasing temperature, $\alpha$ becomes smaller at high fields. The occurrence of linear behavior at $T$ < 50 K may reflect that the system is at the quantum limit [35-38], though the oscillations in the magnetoresistivity may originate from other mechanisms besides Shubnikov−de Haas effect [35,36,38] (see Fig.S3 for additional experimental results and its caption for more discussion). Figure 1(b) shows the temperature dependence of the longitudinal resistivity $\rho_{xx}(T)$ at a few fixed magnetic fields, which are constructed from the measured $\rho_{xx}(H)$ curves at fixed temperatures to avoid nonequilibrium temperature effects. At zero field, the sample shows the expected metallic behavior with a residual resistivity ratio $rrr \approx 14.3$. Similar to those reported in Ref.31, the curves at $H = 0.04$ T, $0.08$ T, and $0.2$ T exhibit the typical 'turn-on' temperature behavior often observed in topological semimetals, i.e., $\rho_{xx}(T)$ changes from metallic to semiconducting-like behavior with decreasing temperature [31]. At $H \geq 0.6$ T, the curves show pure semiconducting-like behavior, which is barely seen in both topological and trivial semimetals. Another notable finding is the magnetoresistance at room temperature, with a remarkable value of $MR \approx 10^4$ % at $H = 9$ T (see Fig.2(a) and Fig.2(d)). As shown in Figs.2(b) and 2(e), the MRs also violate the Kohler's rule.

It is challenging to determine carrier densities in semimetals using transport measurements. Typically, they are derived by fitting the magnetoresistivities/magnetoconductivities with a two-band model [18-20,23-27]. This approach can be unreliable, since most semimetals have multiple bands. For example, the magnetoresistance of ZrSiSe was found to follow Kohler's rule [39] that is valid for systems with constant carrier densities, even though the existence of a Lifshitz transition in this material is inferred from anomalies in $n(T)$ using a two-band model analysis [18].

Graphite is a two-band system with one electron-pocket and two hole-pockets [28]. Thus, its magnetoresistivities were expected to follow the two-band model. Indeed, $\rho_{xx}(H)$ curves along with the Hall magentoresisitivity $\rho_{xy}(H)$ curves at high temperatures ($T > 120$ K) can be described with the two-band model, as shown in Fig.S4(a) for those at $T = 300$ K. However, deviations between the fitting and experimental curves are noticeable. In fact, $\rho_{xx} \sim H^{1.8}$ relationship shown in the inset of Fig.S2 for the experimental $\rho_{xx}(H)$ curve obtained at $T = 300$ K differs from that ($\rho_{xx} \sim H^2$) expected for a compensated two-band system. The resulting carrier densities $n_e$ and $n_h$, though being close to the theoretical values, have weaker temperature dependences than the predicted one (Fig.S4(c)). The reason can be ascribed to the distribution of the effective mass, resulting in a magnetic field dependent exponent $\alpha$ in the $\rho_{xx} \sim H^\alpha$ relationship [28]. At $T < 120$ K the deviations between the fitting and experimental curves become more pronounced with decreasing temperature, as shown in Fig.S4(b) for those at $T = 104$ K. For $T < 70$ K, the two-band model fails completely, probably due to quantum effects [35-38]. That is, fittings of the magnetoresistivity data with the two-band model is not a reliable method to quantitively determine the carrier density and mobility in graphite.

On the other hand, the extended Kohler's rule Eq. (1) can reveal the temperature dependence of the carrier density if the densities and mobilities in different bands have the same/similar temperature dependences [33], which is the case in graphite as shown in Fig.S4. We find that both $\rho_{xx}(H)$ and $\rho_{xx}(T)$ obey Eq.1 (see Fig.S5), i.e., $MR = f[H/(n_T \rho_0)]$ [33], with pronounced temperature dependence of $n_T$. This is not a surprise, since the carrier density in graphite at zero temperature $n_0$ (~$10^{18}$ cm$^{-3}$) is very close to that of TaP, in which the extended Kohler's rule was established [33]. The derived temperature dependence of the carrier density $n_T$ (see Fig.S6) is also similar to that of TaP: it decreases with temperature at $T > 50$ K and saturates at lower temperatures. We find that such a $n_T$ can be quantitatively described theoretically [28], considering the region where the lattice scattering is dominant. The consistency of the experimental results with theory can be seen from the extended Kohler's rule plots of $\rho_{xx}(H)$ and $\rho_{xx}(T)$ using the theoretical value of $n_T$, as presented in Figs.2(c) and 2(f), where $n_T$ is obtained by normalizing the theoretical temperature-dependent carrier densities to that at $T = 300$ K (see Fig.S6). Since $n_T$ was calculated with a temperature-dependent Fermi energy $E_F$ (see inset of Fig.S6), the scaling behavior shown in Figs.2(c) and 2(f) infers possible temperature-induced change in the size of the Fermi pockets.

The above finding led us to probe the temperature effects on the shape of the Fermi surface. Revealing the temperature-driven change in the shape of the Fermi surface can also provide additional support to the claim of temperature-dependent size. The shape of the Fermi surface is reflected in the effective mass $m^*$ of the charge carriers, which governs the magnetoresistivity through the carrier mobility $\mu$ ($\sim 1/m^*$) [32]. Thus, we experimentally determined the anisotropy of $m^*$ by measuring the magnetoresistivity at a given temperature with varying magnetic field orientations $\theta$ (definition is shown in the inset of Fig.3(c)). Figure 3(a) shows $\rho_{xx}(H)$ curves obtained at $T = 300$ K and at various angles $\theta$. They clearly show that the magnetoresistivity is anisotropic, with the smallest values at $H//ab$, i.e., $\theta = 90°$. Following the procedures in Ref.[32], we re-plot the data as $\rho_{xx} \sim \varepsilon_\theta H$ in Fig.3(b) with the resulting $\varepsilon_\theta$ shown in Fig.3(c). The collapse of all curves for $\theta \neq 0°$ onto the curve at $\theta = 0°$ in Fig.3(b) follows the scaling behavior of Eq.2, i.e., $\rho_{xx}(H,\theta) = \rho_{xx}(\varepsilon_\theta H)$, with $\varepsilon_\theta = (cos^2\theta + sin^2\theta/\gamma^2)^{1/2}$ and $\gamma = 20$ as shown in Fig.3(c). The observed anisotropy of the magnetoresistivity and scaling behavior can be further confirmed by measuring the angle dependence of the magnetoresistivity $\rho_{xx}(\theta)$ at a particular magnetic field value $H$. Here, data is taken at a constant magnetic field while rotating the sample with respect to the external magnetic field, as demonstrated by $\rho_{xx}(\theta)$ curves obtained at various fields and their scaling in Figs.S7(a) and S7(b), respectively.

The temperature effects on the shape of the Fermi surface can be inferred from the temperature dependence of $\gamma$, which represents the anisotropy of the effective mass $m^*$ for an ellipsoidal Fermi surface [32]. To reveal $\gamma$'s temperature behavior, we repeated the measurements of $\rho_{xx}(H)$ and $\rho_{xx}(\theta)$ and the associated analysis procedures for $T = 300$ K at other temperatures. As examples, we present $\rho_{xx}(H)$ data and their scaling analysis for $T = 5$ K in Fig.S8 while Fig.S9 shows $\rho_{xx}(\theta)$ data obtained at this temperature and the corresponding scaling. They clearly exhibit the high anisotropy of the magnetoresistivities and excellent scaling with Eq.2. The derived $\gamma$ shows pronounced temperature dependence, with $\gamma = 55$ at $T = 5$ K. The temperature dependence of $\gamma$ is presented in Fig.4 and exhibits a smooth decrease with increasing temperature, ranging from $\gamma = 56$ at $T = 2$ K to $\gamma = 20$ at $T = 300$ K. Interestingly, Figure 4 also shows a strong correlation between the $\gamma$'s temperature dependence and that of the magnetoresistance *MR*, though the overlap of $\gamma(T)$ and $MR(T)$ may be coincidental.

Since the electron and hole bands in graphite have different densities of states, the temperature-driven shift of the Fermi level is caused by the requirement of compensation of $n_e = n_h$ at all temperatures, where $n_e$ and $n_h$ are the electron and hole density, respectively [28]. Pronounced temperature-effects on the Fermi surface occur because the Fermi level is close to the bottom of the conduction band and the top of the valence band, i.e., the sizes of the electron and hole pockets are small. On the other hand, a typical semimetal has more than two-bands and the total density of the electrons and holes at zero temperature may differ from each other [33]. However, the changes in the electron and hole density induced by temperature must be equal, i.e., $\Delta n_e = \Delta n_h$, which can result in a temperature-dependent Fermi level if the density of states of the electron and hole bands are not the same. That is, it may be not uncommon to observe a temperature-induced shift of the Fermi level, i.e., change in the size of the Fermi pockets in semimetals. Lifshitz transition occurs when a shift of the Fermi level eventually leads to a change in the Fermi surface topology, e.g., disappearance and/or emergence of a Fermi pocket [7]. If the carrier density is low, the change in Fermi level can have a significant impact on the temperature dependence of the carrier density. In this case, the size change of the Fermi pockets may be deduced from $n_T$ in the extended Kohler's rule of magnetoresistance.

In probing the role of the quasi-2D band on the occurrence of the extremely large magnetoresistance, a temperature-dependent magnetoresistance anisotropy was first discovered in the type-II Weyl semimetal $WTe_2$ [32]. This phenomenon was later observed in both topological and trivial semimetals [15,32,40-44]. In a semimetal with multiple anisotropic Fermi pockets, such a phenomenon could occur if the temperature dependences of the carrier mobilities of different Fermi pockets are not identical. On the other hand, our results from graphite indicate that such a temperature-dependent magnetoresistance anisotropy can be the direct outcome of a temperature-driven change in the anisotropy of the Fermi surface. That is, we can use temperature-dependent anisotropy of magnetoresistance, which can be obtained through convenient and widely available transport measurements, to search for possible temperature-driven changes in the anisotropy of the Fermi surface in a semimetal.

## IV. CONCLUSIONS

In summary, we probed temperature effects on the Fermi surface of graphite using magnetotransport measurements. Temperature-induced size change of the Fermi pockets is inferred from the temperature dependence of the carrier density derived from the extended Kohler's rule plots of the magnetoresistances, obtained in a given magnetic field orientation and at various temperatures. Temperature-driven change in the anisotropy of the Fermi pockets is revealed from the temperature-dependent anisotropy of the magnetoresistance. Our results show that temperature-induced changes in the Fermi surface may be expected in semimetals. They also indicate that temperature-induced changes in the Fermi surface need to be considered in understanding the temperature behavior of a semimetal, such as the temperature-dependent carrier density and the anisotropy of the magnetoresistances. This work further demonstrates that the widely available magnetotransport measurements can be used to detect temperature effects on the Fermi surface of a semimetal through the temperature dependence of the magnetoresistance anisotropy and the carrier density, with the latter being highly relevant in systems with low carrier density.

**Acknowledgements**

This project was supported by the U.S. Department of Energy, Office of Science, Basic Energy Sciences, Materials Sciences and Engineering. S. E. P and Z. -L. X. acknowledge support by National Science Foundation (Grant No. DMR-1901843). T. T. W and Y. L. W. received supports by the Jiangsu Excellent Young Scholar program (BK20200008). Work performed at the Center for Nanoscale Materials, a U.S. Department of Energy Office of Science User Facility, was supported by the U.S. DOE, Office of Basic Energy Sciences, under Contract No. DE-AC02-06CH11357.

**Figure captions**

**FIG.1.** (color online) Magnetoresistivities of graphite. (a) Magnetic field dependence $\rho_{xx}(H)$ measured at various temperatures. For clarity we plot only a portion of the $\rho_{xx}(H)$ curves taken at temperatures from $T = 2$ K to 120 K at intervals of 2 K; 123K to 180 K at intervals of 3K; 184 K to 200 K at intervals of 4 K, and from 205 K to 300 K at intervals of 5 K. (b) Temperature dependence $\rho_{xx}(T)$ constructed from $\rho_{xx}(H)$ data. For clarity we present $\rho_{xx}(T)$ curves at magnetic fields of $H$ = 0 T, 0.02 T, 0.04 T, 0.08 T, 0.2 T, 0.6 T, 1.4 T, 2.6 T, 4.2 T, 6.2 T and 9 T (from bottom to top). The data were taken in magnetic field parallel to the *c*-axis of the crystal.

**FIG.2.** (color online) Extended Kohler's rule of the magnetoresistance. (a) and (d), Magnetic field and temperature dependences of the *MR* derived from data in Figs.1(a) and 1(b), respectively. (b) and (e), Kohler's rule plots of the data in (a) and (d), respectively. (c) and (f), Extended Kohler's rule plots of the *MR* curves in (a) and (d), respectively. To be comparable with the scaling results in Fig.S6, theoretical values of $n_T$ were calculated by normalizing the temperature-dependent carrier densities in Fig.S6 to that at $T = 300$ K, i.e., $n_T = 1$ at $T = 300$ K. Symbols in Fig.2(a)-2(c) are the same as those in Fig.1(a) while the same symbols as those in Fig.1(b) are used for Fig.2(d)-2(f) (see legends in Figs.2(c) and 2(f), respectively). Dashed red lines in Fig.2(c) and 2(f) represent a power-law relationship of $MR \sim H^\alpha$ with $\alpha = 1.8$.

**FIG.3.** (color online) Anisotropy of the magnetoresistivity in graphite. (a) $\rho_{xx}(H)$ curves at various angles $\theta$ obtained at $T = 300$ K. (b) Data in (a) re-plotted with $H$ scaled by a factor $\varepsilon_\theta$. (c) Angle dependence of the scaling factor $\varepsilon_\theta$. Symbols are experimental data and the solid line is a fit with $\varepsilon_\theta = (cos^2\theta + sin^2\theta/\gamma^2)^{1/2}$ and $\gamma = 20$. A schematic in the inset of (c) shows the definition of angle $\theta$.

**FIG.4.** (color online) Temperature dependences of the derived anisotropy factor $\gamma$ (red open circles) and the magnetoresistance $MR$ at $H = 0.2$ T (solid line).

**Figure 1**

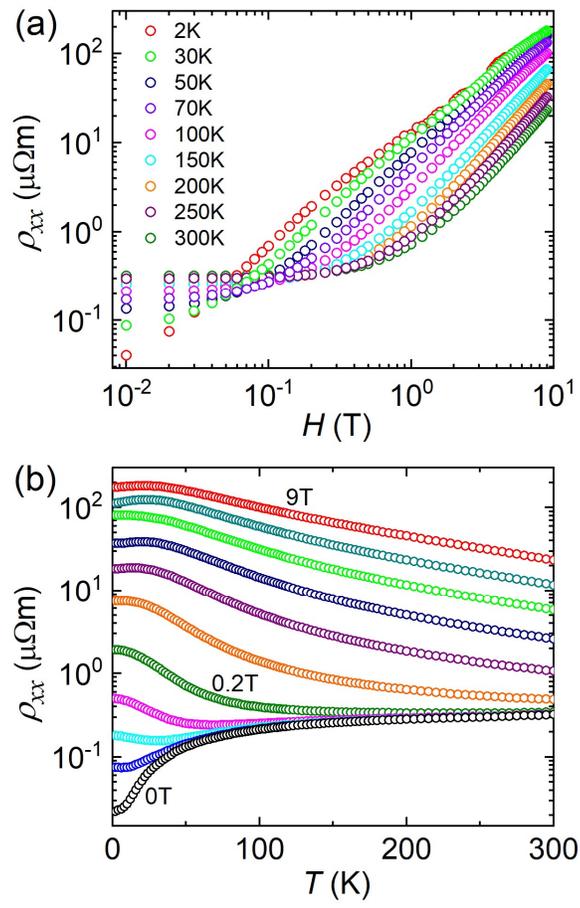

**Figure 2**

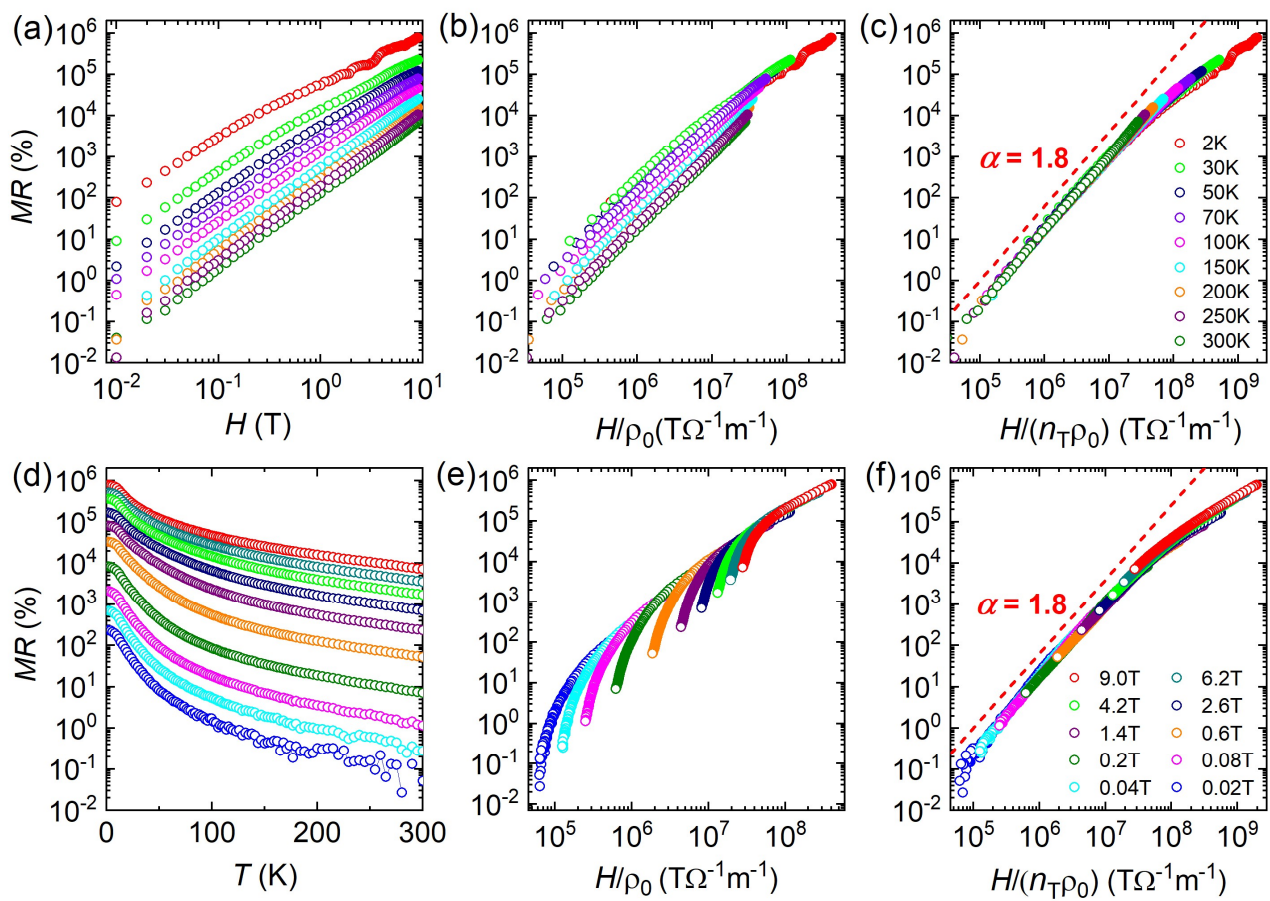
17

**Figure 3**

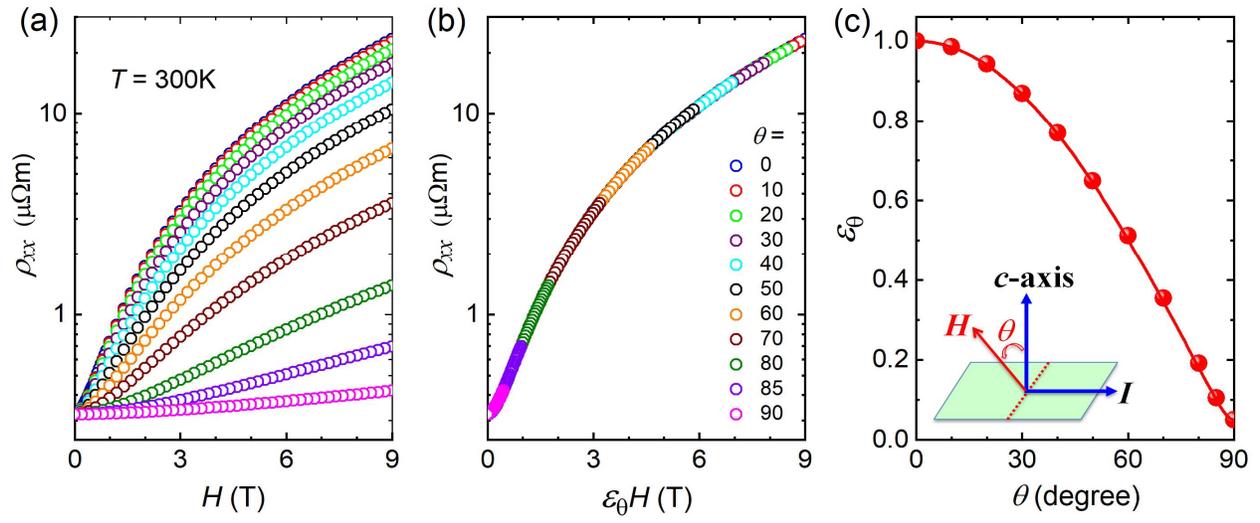

**Figure 4**

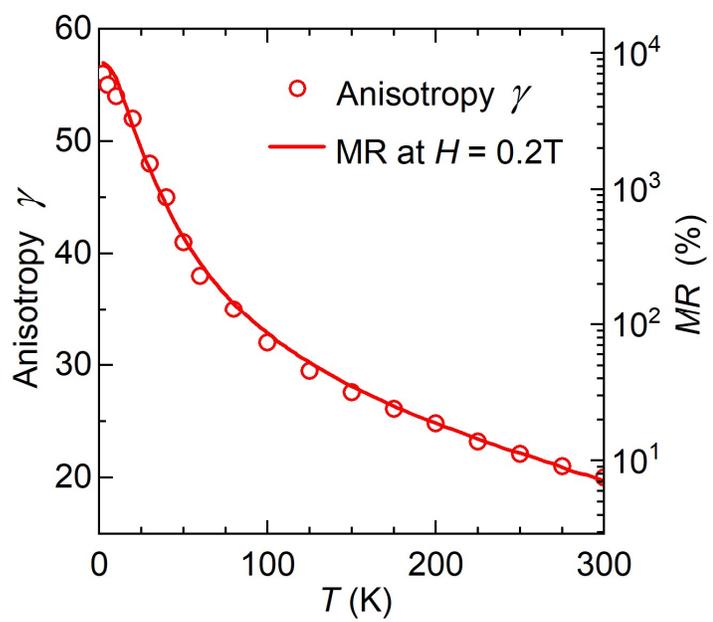

# Temperature-driven changes in the Fermi surface of graphite

Laxman R. Thoutam[1,2], Samuel E. Pate[1,3], Tingting Wang[4], Yong-Lei Wang[4,*], Ralu Divan[5], Ivar Martin[1], Adina Luican-Mayer[6], Ulrich Welp[1], Wai-Kwong Kwok[1], and Zhi-Li Xiao[1,3,*]

[1]*Materials Science Division, Argonne National Laboratory, Argonne, Illinois 60439, USA*

[2]*Department of Electronics and Communications Engineering, SR University, Warangal, Telangana 506371, India*

[3]*Department of Physics, Northern Illinois University, DeKalb, Illinois 60115, USA*

[4]*Research Institute of Superconductor Electronics, School of Electronic Science and Engineering, Nanjing University, Nanjing 210093, China*

[5]*Center for Nanoscale Materials, Argonne National Laboratory, Argonne, Illinois 60439, USA*

[6]*Department of Physics, University of Ottawa, Ottawa, Ontario K1N 6N5, Canada*

[*]Correspondence to: yongleiwang@nju.edu.cn; xiao@anl.gov

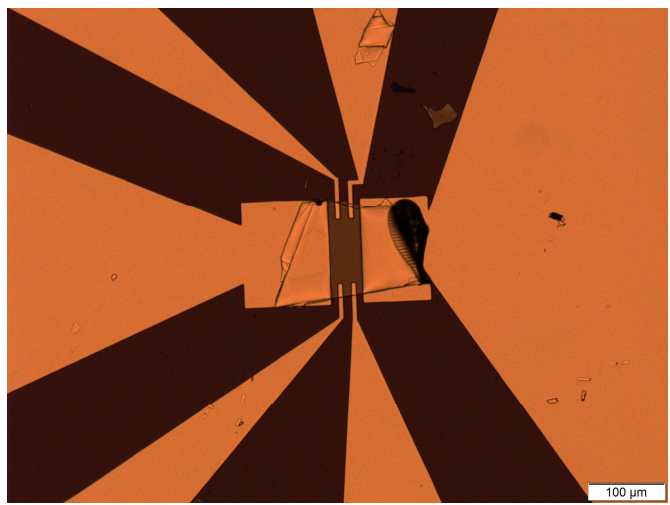

**Fig.S1.** Optic image of an exfoliated graphite flake with current and voltage leads. Its thickness determined using atomic force microscopy is $d$ = 400 nm. The width and separation between the voltage leads long the current direction are $w$ = 120 μm and $l$ = 18 μm, respectively. The current flows from left to right.

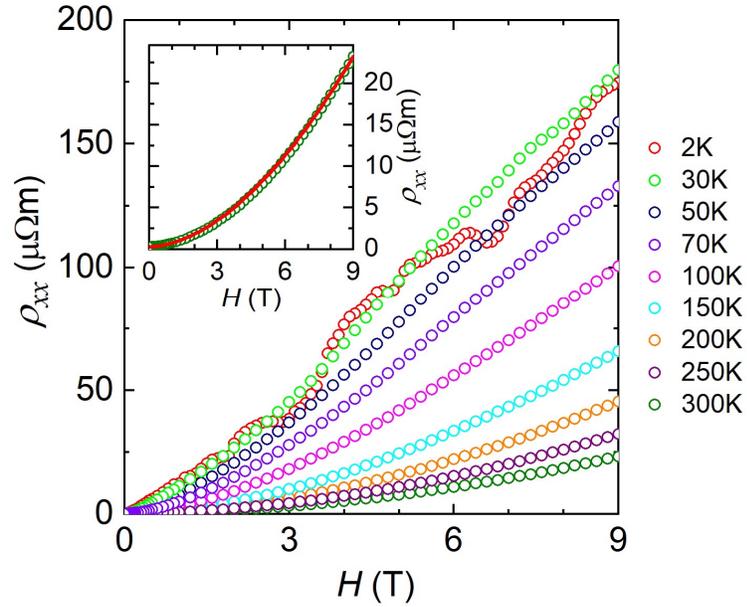

**Fig.S2.** Magnetic field dependence of the resistivity $\rho_{xx}(H)$ measured at various temperatures. The data are the same as those in Fig.1(a) but plotted in a linear scale to show the change in curvature at high magnetic fields. The inset illustrates the data at $T =$ 300 K, which follows the relationship $\rho_{xx} \sim H^\alpha$ with $\alpha = 1.8$ (red line).

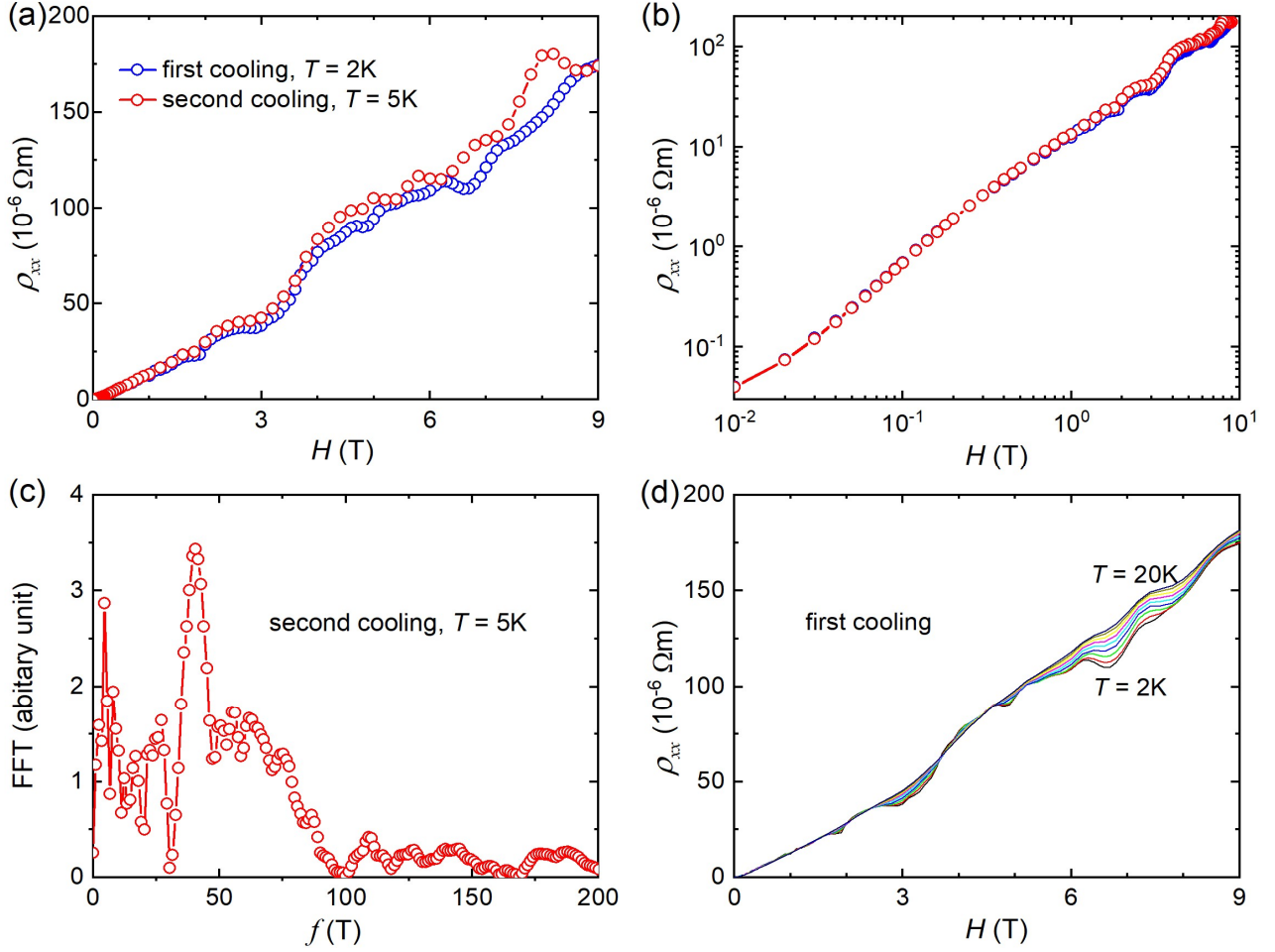

**Fig.S3.** Cooling history dependence of the magnetoresistance oscillations. (a) Comparison of $\rho_{xx}(H)$ curves obtained from two cooling runs. In experiments, we first cooled the sample to $T = 2$ K from $T = 300$ K and obtained all $R(H)$ curves at H//c ($\theta = 0°$) starting at $T = 2$ K and ending at $T = 300$ K, which yielded the data in Fig.1 for the extended Kohler's rule analysis. Subquentially, we cooled the sample down to 2 K again to obtain $R(H)$ curves at various angles and $R(\theta)$ curves at various magnetic fields at different temperatures, which yielded data as those in Fig.3a and Figs.S7-S9 for the anisotropy analysis. Difference in oscillation patterns can be clearly identified. (b) Data of (a) in logarithmic scale, showing overlap of the two curves at low magnetic field ($H < 1$ T). (c) Fast Fourier transform (FFT) analysis of the MR oscillation in $\rho_{xx}(H)$ curve at $T = 5$ K. We could not reliably identify two peaks expected from the electron and hole pockets, probably due to limited numbers of oscillation periods in the narrow magnetic field range. (d) $\rho_{xx}(H)$ curves from $T = 2$ K to 20 K at intervals of 2 K obtained during first cooling, showing that the oscillation patterns are temperature-independent in the same cooling run.

The difference in oscillation patterns obtained in different cooling runs indicates that the MR oscillations in graphite can depend on the cooling history. MR oscillations besides those from the Shubnikov-de-Haas (SdH) effect have been reported in graphite in the quantum limit and various origins have been proposed [35,36,38], e.g., quantum Hall effect or superconducting correlations [35,36], and the existence of moiré superlattices in the layered graphite [38]. The cooling history effect of the MR oscillations seen here tends to favor an extrinsic origin such as moiré superlattices that could differ under each cooling run due to uncontrollable structural change/strain.

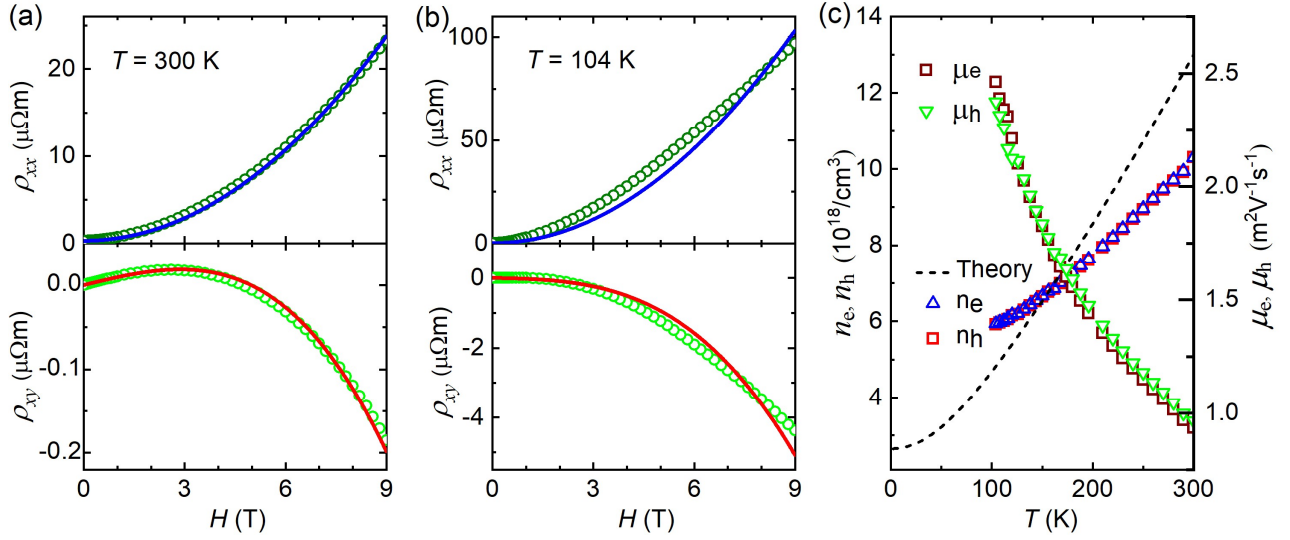

**Fig.S4.** Two-band model analysis of the magnetoresistivities $\rho_{xx}(H)$ and $\rho_{xy}(H)$ at $T = 300$ K (a) and at $T = 104$ K (b). (c) The derived carrier densities ($n_e$ and $n_h$) and mobilities ($\mu_e$ and $\mu_h$). The dashed line in (c) represents the theoretical values of the carrier density, where $n_e = n_h$ is assumed [28]. Both $\rho_{xx}(H)$ and $\rho_{xy}(H)$ at the same temperature are fit simultaneously with the following equations [33] from the two-band model:

$$\rho_{xx}(H) = \frac{1}{e}\frac{(n_h\mu_h + n_e\mu_e) + (n_h\mu_e + n_e\mu_h)\mu_h\mu_e H^2}{(n_h\mu_h + n_e\mu_e)^2 + (n_h - n_e)^2\mu_h^2\mu_e^2 H^2}$$

$$\rho_{xy}(H) = \frac{H}{e}\frac{(n_h\mu_h^2 - n_e\mu_e^2) + (n_h - n_e)\mu_h^2\mu_e^2 H^2}{(n_h\mu_h + n_e\mu_e)^2 + (n_h - n_e)^2\mu_h^2\mu_e^2 H^2}$$

At low temperatures the two-band model fails due to quantum effects [35-38]. No anomalies in both the carrier density and mobility imply the absence of Lifshitz transition in graphite in this temperature regime. $n_e \approx n_h$ and $\mu_e \approx \mu_h$ also enable the extended Kohler's rule Eq.1 with the thermal factor $n_T \sim n_e(T)$ (or $n_h(T)$), since $n_T = [\sum_i(n_i\mu_i)]^{3/2}/[\sum_i(n_i\mu_i^3)]^{1/2}$ in systems with two or more bands, i.e., $i \geq 2$ [Ref.33].

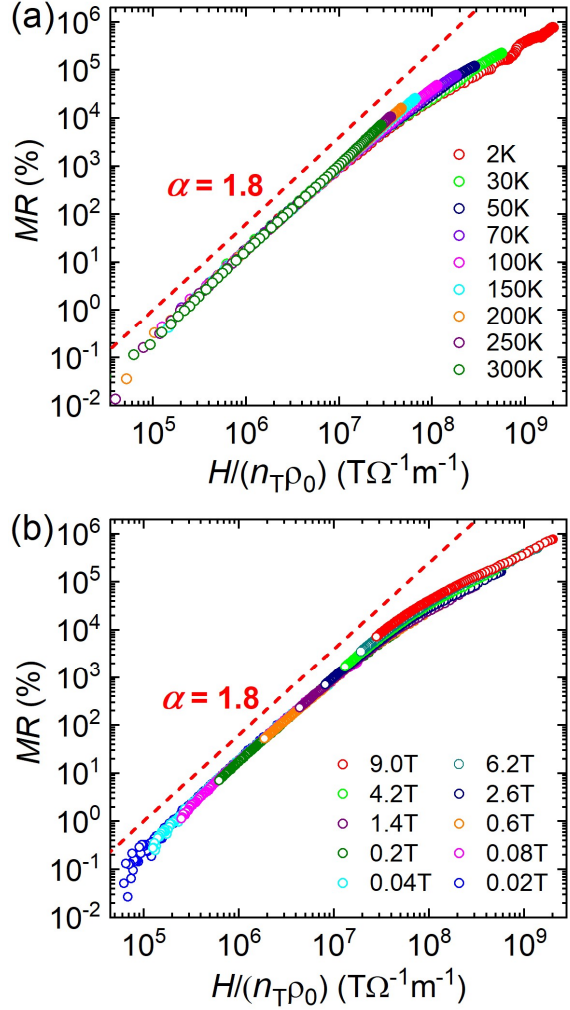

**Fig.S5.** Extended Kohler's rule plots of the magnetoresistivity. (a) and (b) correspond to the data in Fig.1(a) (Fig.2(a)) and Fig.1(b) (Fig.2(d)), respectively.

We followed the scaling procedures described in Ref.[33]: we first prepared the Kohler's rule plots $MR \sim H/\rho_0$ (Fig.2(b) and Fig.2(e)), and then multiplied a temperature factor of $1/n_T$ to $H/\rho_0$ (x-axis) in each of the $MR \sim H/\rho_0$ curves. For simplification, we assumed $n_T = 1$ for the curve at $T = 300$ K. That is, we scaled all the $MR$ curves to the $T = 300$ K curve by varying $n_T$ for each curve. The derived $n_T$ values at various temperatures are presented in Fig.S6. It decreases from $n_T = 1$ at $T = 300$ K to $n_T \approx 0.2$ at $T = 2$ K.

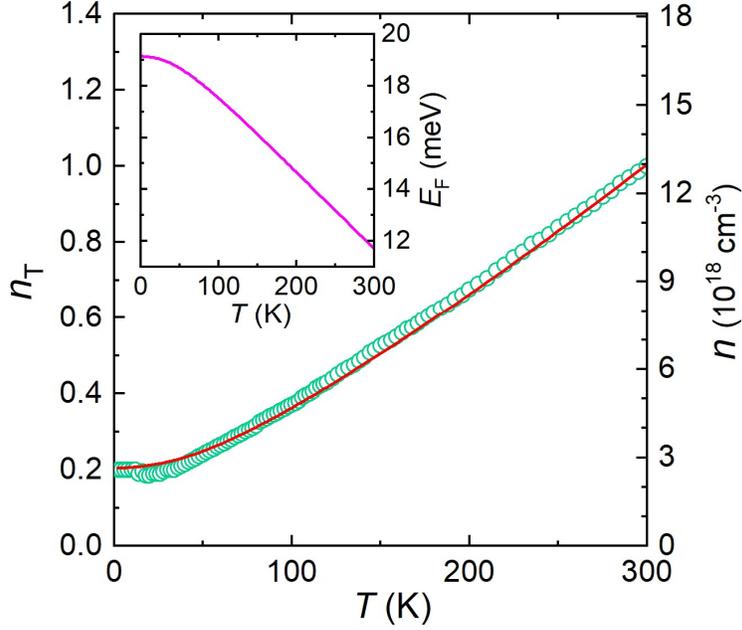

**Fig.S6.** Comparison of the temperature dependence of the experimental carrier density $n_T$ (green circles) and the calculated carrier density $n$ (red line). $n_T$ is the temperature factor in the extended Kohler's rule plot and its value is set to be 1, i.e., $n_T = 1$, at $T = 300$ K (see Fig.S4 and its caption). In theory [28], the densities of electrons or holes in the compensated graphite are assumed to be equal, i.e., $n_e = n_h = n$ at all temperatures, resulting in a temperature dependent Fermi energy $E_F$ presented in the inset. The implemented equations and material parameters are [28]:

$$n_e = \frac{4}{c_0\pi^2\eta_0^2}\int_0^{\pi/2} d\varphi \frac{1}{(1+v)^2}\int_{E_3}^{\infty} dE f_e(E)(E - \frac{E_2 + E_3}{2})$$

$$n_h = \frac{4}{c_0\pi^2\eta_0^2}\int_0^{\varphi_0} d\varphi \frac{1}{(1-v)^2}\int_{-\infty}^{E_3} dE f_h(E)(\frac{E_1 + E_3}{2} - E)$$

where

$$f_e(E) = 1/[1 + \exp\left(\frac{E - E_F}{k_B T}\right)] \text{ and } f_h(E) = 1 - f_e(E)$$

$$E_1 = \Delta + 2\gamma_1 \cos\varphi + 2\gamma_5 \cos^2\varphi$$

$$E_2 = \Delta - 2\gamma_1 \cos\varphi + 2\gamma_5 \cos^2\varphi$$

$$E_3 = 2\gamma_2 \cos^2\varphi$$

$$\varphi = \tfrac{1}{2}k_z c_0;\ c_0 = 6.71\text{Å};\ \eta_0 = \tfrac{\sqrt{3}}{2}\gamma_0 a_0;\ a_0 = 2.46\text{Å};\ v = \tfrac{2\gamma_4}{\gamma_0}\cos\varphi$$

$$\Delta = -0.02;\ \gamma_0 = 2.88;\ \gamma_1 = 0.395;\ \gamma_2 = 0.016;\ \gamma_4 = -0.20;\ \gamma_5 = 0.016$$

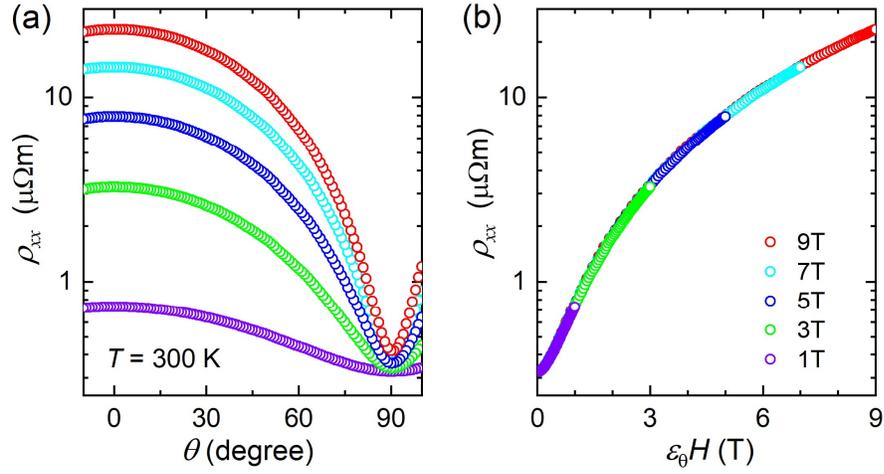

**Fig.S7.** (a) Angle dependence of the magnetoresistivity $\rho_{xx}(\theta)$ at $T = 300$ K and at various magnetic fields. (b) Data in (a) re-plotted with $H$ scaled by a factor $\varepsilon_\theta = (cos^2\theta + sin^2\theta/\gamma^2)^{1/2}$ and $\gamma = 20$, is consistent with results in Fig.3 on $\rho_{xx}(H)$ curves at various angles.

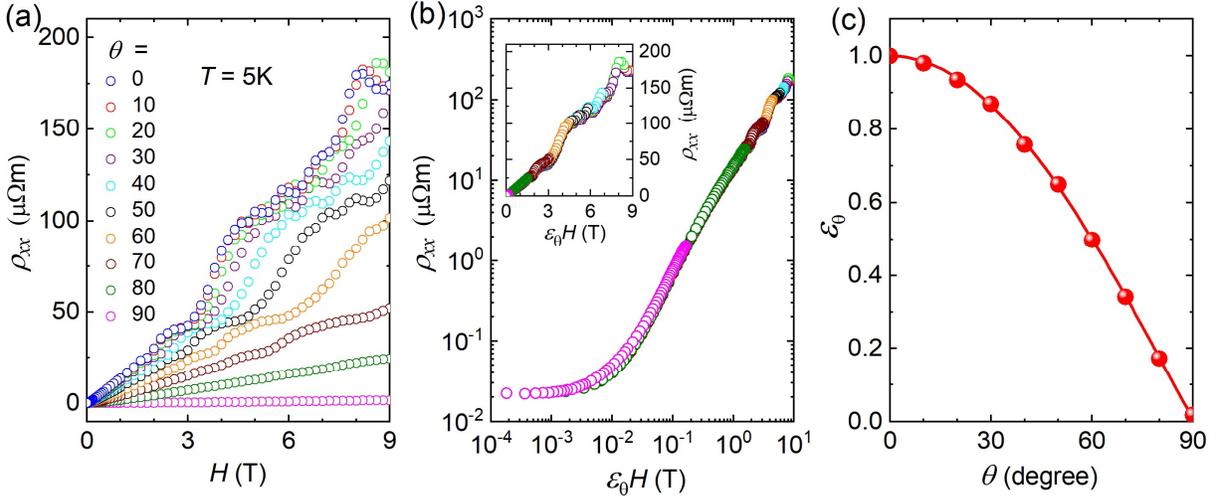

**Fig.S8.** Anisotropy of the magnetoresistivity. (a) $\rho_{xx}(H)$ curves at various angles $\theta$ obtained at $T = 5$ K. (b) Data in (a) re-plotted with $H$ scaled by a factor $\varepsilon_\theta$. (c) Angle dependence of the scaling factor $\varepsilon_\theta$. Symbols are experimental data and the solid line is a fit with $\varepsilon_\theta = (cos^2\theta + sin^2\theta/\gamma^2)^{1/2}$ and $\gamma = 55$.

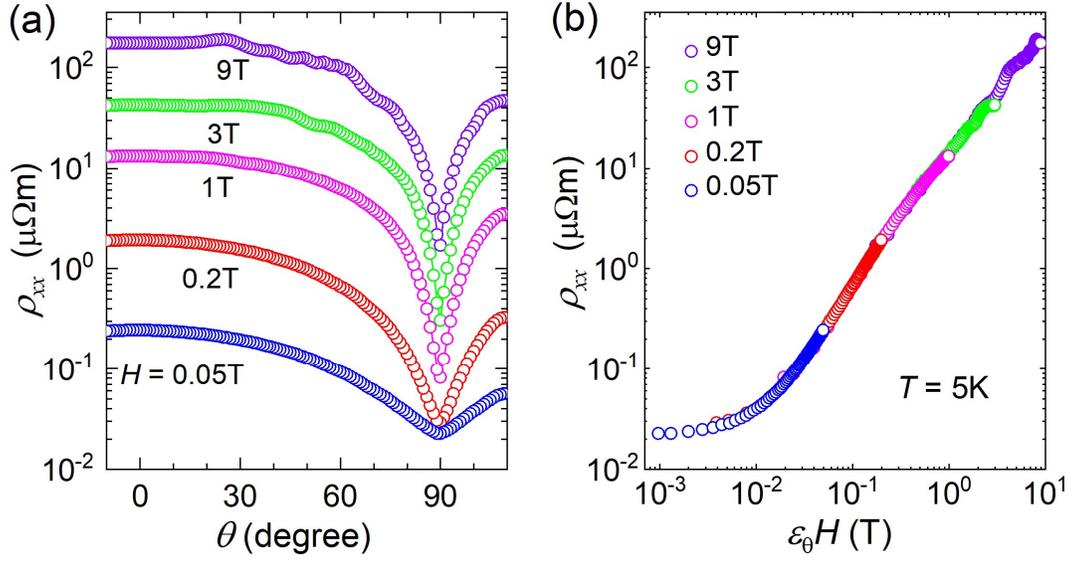

**Fig.S9.** (a) Angle dependence of the magnetoresistivity $\rho_{xx}(\theta)$ at $T$ = 5 K and various magnetic fields. (b) Data in (a) re-plotted with $H$ scaled by a factor $\varepsilon_\theta = (cos^2\theta + sin^2\theta/\gamma^2)^{1/2}$ and $\gamma = 55$, is consistent with results in Fig.S8 on $\rho_{xx}(H)$ curves at various angles.